\documentclass[pra,reprint,superscriptaddress,amsmath,amssymb,aps,floatfix]{revtex4-1}

\usepackage{hyperref}
\usepackage{graphicx}

\usepackage[notext]{stix}
\usepackage{times}

\DeclareSymbolFont{CMAlt}{OMX}{cmex}{m}{n}
\DeclareMathSymbol{\sumop}{\mathop}{CMAlt}{"50}

\usepackage{color}
\hypersetup{%
    unicode=true,
    colorlinks=true,
    linkcolor=blue,
    citecolor=blue,
    urlcolor=blue,
    pdftitle={Orbital ordering of ultracold alkaline-earth atoms in optical lattices},
    pdfauthor={A. Sotnikov, N. Darkwah Oppong, Y. Zambrano, A. Cichy},
    pdfproducer={},
    pdfcreator={}}

\newcommand{\asciimathunit}[1]{\ensuremath{\,\mathrm{#1}}}
\newcommand{\nm}{\asciimathunit{nm}}

\newcommand{\subfigref}[2]{\hyperref[fig:#1]{\ref*{fig:#1}(#2)}}
\newcommand{\subfigrefs}[3]{\subfigref{#1}{#2}--\subfigref{#1}{#3}}

\begin{document}
\title{Orbital ordering of ultracold alkaline-earth atoms in optical lattices}

\author{Andrii Sotnikov}
\email{a\_sotnikov@kipt.kharkov.ua}
\affiliation{Akhiezer Institute for Theoretical Physics, NSC KIPT, Akademichna 1, 61108 Kharkiv, Ukraine}
\affiliation{Karazin Kharkiv National University, Svobody Sq. 4, 61022 Kharkiv, Ukraine}

\author{Nelson \surname{Darkwah Oppong}}
\affiliation{Max-Planck-Institut f{\"u}r Quantenoptik, Hans-Kopfermann-Stra{\ss}e 1, 85748 Garching, Germany}
\affiliation{Munich Center for Quantum Science and Technology (MCQST), Schellingstra{\ss}e 4, 80799 M{\"u}nchen, Germany}
\affiliation{Ludwig-Maximilians-Universit{\"a}t, Schellingstra{\ss}e 4, 80799 M{\"u}nchen, Germany}

\author{Yeimer Zambrano}
\author{Agnieszka Cichy}
\affiliation{Faculty of Physics, Adam Mickiewicz University, Uniwersytetu Pozna\'nskiego 2, 61-614 Poznan, Poland}

\date{\today}

\begin{abstract}
We report on a dynamical mean-field theoretical analysis of emerging low-temperature phases in multicomponent gases of fermionic alkaline-earth(-like) atoms in state-dependent optical lattices.
Using the example of $^{173}$Yb atoms, we show that a two-orbital mixture with two nuclear spin components is a promising candidate for studies of not only magnetic but also staggered orbital ordering peculiar to certain solid-state materials.
We calculate and study the phase diagram of the full Hamiltonian with parameters similar to existing experiments and reveal an antiferro-orbital phase.
This long-range-ordered phase is inherently stable, and we analyze the change of local and global observables across the corresponding transition lines, paving the way for experimental observations.
Furthermore, we suggest a realistic extension of the system to include and probe a Jahn-Teller source field playing one of the key roles in real crystals.
\end{abstract}

\maketitle
\linepenalty=1000

\section{Introduction}
In solid-state materials, electrons can occupy different orbital states, which usually determine their directional mobility.
Besides spin and charge, this orbital degree of freedom plays an important role in interacting electron systems and lies at the heart of intriguing many-body phenomena such as colossal magnetoresistance, heavy fermions, and the Kondo effect~\cite{Moritomo1996,Hewson1997,Tokura00}.

In particular, orbital ordering is one of the key phenomena in materials with multiorbital structure such as transition-metal oxides.
Similar to the ordered pattern of spins in the ground state of an antiferromagnet, electrons from different \mbox{$d$-orbital} manifolds can spatially arrange in these materials~\cite{Khomskii2014}.
While great advances have been made in both the experimental observation and the theoretical description of orbital ordering~\cite{Khalifah02,Keimer06,Miller2015,Ishigaki19}, the microscopic origin of the processes is still not fully understood.
Numerical calculations for these systems are challenging due to the simultaneous presence of electron-electron as well as electron-phonon interactions, which both can effect orbital ordering~\cite{Pavarini2012,Pavarini2019}.
Therefore, quantum simulations of the corresponding model Hamiltonians could shed light onto the competing mechanisms and the nature of orbitally-ordered phases.

Ultracold atoms have become a versatile and clean platform for quantum simulations of solid-state systems in the last decade~\cite{Gross17}.
In particular, ultracold fermionic atoms in optical lattices have allowed to study the single-orbital Fermi-Hubbard model~\cite{Har2015Nat,Cheuk16,Boll16,Parsons16}, which is believed to describe certain high-temperature superconductors~\cite{Micnas90,Keimer2015,Fradkin2015}.
More recently, a two-orbital Fermi-Hubbard system has been realized with ultracold alkaline-earth(-like) atoms (AEAs) in a state-dependent optical lattice (SDL)~\cite{Riegger18}.
For atoms of this kind, the availability of the long-lived metastable $^3$P$_0$ electronic state (denoted as $e$) in addition to the $^1$S$_0$ ground state (denoted as $g$) allows populating two orbital states of the lattice with distinct kinetic and interaction properties, which makes these atoms attractive candidates for the study of orbital phenomena~\cite{Gorshkov2010NP,FossFeig10,FossFeig10a,Nonne2011,SilvaValencia2012,Kuzmenko2016,KanaszNagy2018,Nakagawa2018,Goto2019,Zhang2020}.

In this work, we report on the possibility to approach and simulate orbital ordering with AEAs in SDLs.
Our study is based on dynamical mean-field theory (DMFT) applied to a two-orbital Fermi-Hubbard model with parameters closely related to existing experimental implementations with $^{173}$Yb atoms~\cite{Riegger18}.
Nevertheless, our results are also applicable to other fermionic AEAs due to the similarity of relevant interaction properties~\cite{Gorshkov2010NP}.
We calculate the phase diagram for realistic orbital fillings and, in addition to multiple magnetically-ordered phases, we find a particular stable orbitally-ordered phase.
The ordering instability in this system is driven by the different intra-~and interorbital onsite interaction of atoms in the $g$ and $e$ state, corresponding to electron-electron interactions in a solid state system.
Transitions to this long-range-ordered phase result in noticeable changes of experimentally-accessible observables, which we determine for the fraction of doubly-occupied lattice sites, the orbital density distribution in a harmonic trap, and nearest-neighbor correlations.
We also show how the influence of electron-phonon interactions in the form of the Jahn-Teller effect (JTE) can be probed with a suitable superlattice potential or by adjusting the density of atoms appropriately.

\section{System, Model, and Method}\label{sec:system-model-method}
\begin{figure}
  \includegraphics[width=\linewidth]{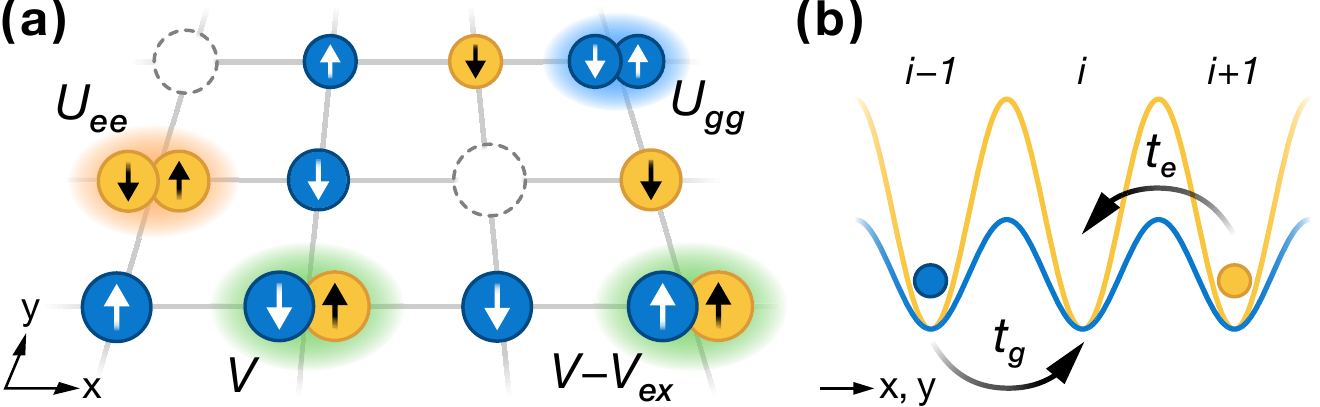}
  \caption{\label{fig:schematic}%
    (a)~Schematic representation of an exemplary  state from the Hilbert space of Eq.~\eqref{eq:hamiltonian} illustrating the interaction parameters.
    The lattice bonds are shown as gray lines and the blue (yellow) circles refer to $g$ ($e$) atoms in different spin states as indicated by the arrows.
    (b)~Illustration of the state-dependent lattice potential (solid lines) which introduces distinct hopping amplitudes $t_g > t_e$ for $g$ and $e$ atoms (blue and yellow circles).}
\end{figure}

We consider a two-orbital mixture of AEAs in the lowest-energy band of a square optical lattice [see Fig.~\subfigref{schematic}{a}], which is state dependent, i.e., the lattice depth differs for the two orbital states $g$ and $e$.
In addition to the orbital degree of freedom, atoms can occupy one of two nuclear spin states (denoted by $\downarrow$ and $\uparrow$), which are equally populated.
Within the tight-binding approximation, this system can be described by the following two-orbital Hubbard model~\cite{FossFeig10,Gorshkov2010NP}:
\begin{eqnarray}\label{eq:hamiltonian}
    &&\mathcal{H} = 
      -\sum_{{\left\langle i,j \right\rangle},\gamma,\sigma}
      {t}_\gamma c_{{i}\gamma\sigma}^{\dag}c_{{j} \gamma \sigma}^{} +
      \sum_{i,\gamma,\sigma} \mu_{\gamma} n_{i \gamma \sigma}
    \nonumber\\
    &&\quad
      +  \sum_{i,\gamma,\sigma < \sigma'} U_{\gamma\gamma} n_{i \gamma \sigma}  n_{i \gamma \sigma'}
      + V \sum_{i,\gamma\neq\gamma',\sigma<\sigma '} n_{i \gamma \sigma} n_{i \gamma' \sigma'}
    \nonumber\\
    &&\quad
      + \left(V-V_{\rm ex}\right) \sum_{i,\gamma < \gamma',\sigma} n_{i \gamma \sigma} n_{i \gamma' \sigma} \nonumber\\
    &&\quad
      + V_{\rm ex} \sum_{i, \gamma \neq \gamma',\sigma < \sigma'} c_{i \gamma \sigma}^{\dag} c_{i \gamma' \sigma'}^{\dag} 
    c_{i \gamma \sigma'}^{} c_{i \gamma' \sigma}^{}\,,
\end{eqnarray}
where $\left\langle i, j \right\rangle$ denotes the set of nearest-neighbor lattice sites, $c_{i\gamma\sigma}^{\dag}$ ($c_{i\gamma\sigma}^{}$) is the fermionic creation (annihilation) operator of an atom in the orbital $\gamma\in{\{g,e\}}$ with spin $\sigma\in{\{\uparrow,\downarrow\}}$, and $n_{i \gamma \sigma}=c_{i\gamma\sigma}^{\dag} c_{i\gamma\sigma}^{}$ is the corresponding density operator.
Importantly, $\mathcal{H}$ preserves the atomic densities $n_\gamma=n_{\gamma\uparrow} + n_{\gamma\downarrow}$ of each orbital averaged over all lattice sites.
Among the relevant Hubbard parameters are the orbital hopping amplitudes $t_\gamma$, the intraorbital interactions $U_{\gamma\gamma}$ as well as the interorbital  direct interaction~$V$ and the exchange interaction~$V_{\rm ex}$.
We obtain these parameters from the experimentally determined $s$-wave scattering lengths and a band-structure calculation.
In our theoretical approach, the average densities $n_g$ and $n_e$ of atoms in the corresponding orbital state can be freely tuned by adjusting the chemical potentials $\mu_g$ and $\mu_e$.
However, we restrict our study to the regimes of low lattice fillings, $n=(n_g+n_e)\leq~2$, to avoid three-body losses~\cite{Hofrichter16PRX}.

In the following, we focus on parameters for a realistic $^{173}$Yb system with a state-dependent optical lattice, which localizes $e$ atoms, i.e., $t_g > t_e$ as depicted in Fig.~\subfigref{schematic}{b}.
This ensures lossy collisions between $e$ atoms are strongly suppressed~\cite{Gorshkov2010NP,Riegger18,Sponselee19}.
We consider a combination of a square state-dependent lattice and a sufficiently deep perpendicular state-independent lattice (see Appendix~\ref{apx:implementation}), which ensures the system is quasi-two-dimensional (quasi-2D) and can be described by the Hamiltonian in Eq.~\eqref{eq:hamiltonian}.
We note that our theoretical calculation can be directly extended to a three-dimensional system~\cite{Golubeva2015PRA}, but the 2D geometry has advantages for experimental realizations, especially for quantum gas microscopy~\cite{Gross17}.
In general, $g$ and $e$ atoms experience different lattice potentials due to distinct polarizabilities ${\alpha_{g}({\lambda})\neq\alpha_{e}(\lambda)}$ of the respective states at a given wavelength~$\lambda$.
At fixed depth of the SDL, the ratio of orbital mobility, $t_g/t_e$, can be tuned by adjusting this wavelength appropriately.
The associated polarizability ratio $p=\alpha_{e}(\lambda)/\alpha_{g}(\lambda)$ increases monotonously between certain atomic transition wavelengths, in particular, from $1$ at the so-called magic wavelength ($\lambda=759\nm$) to $3.3$ at $\lambda=670\nm$~\cite{Riegger18,dzuba18}.
We consider this experimentally accessible regime of $p$ to tune the orbital-dependent mobility, which enhances or suppresses ordered phases.

The onsite interaction energies of $^{173}$Yb atoms in the proposed setup demonstrate the hierarchy $U_{gg}<U_{ee}\sim V_{\rm ex}<V$ due to the relatively large scattering length of the orbitally-symmetric state, which contributes both to $V$ and $V_{\rm ex}$~\cite{Gorshkov2010NP,Scazza2014NP,Cappellini2014,Hoefer15,Pagano2015}.
This contrasts with the well-known Slater-Kanamori parametrization of the Coulomb interaction in correlated electron systems~\cite{Kanamori63}, where the largest quantity is the Coulomb parameter $U=U_{gg}=U_{ee}$, and the Hund's coupling is bounded from above, $V_{\rm ex}\leq U/3$, which ensures repulsive interactions for all spin and orbital components.
In the case of cubic symmetry in the orbital space, the direct interaction amplitude is given as $V=U-2V_{\rm ex}$, such that a different hierarchy is observed, $V_{\rm ex}\leq V<U$.
Nevertheless, the relatively small $U_{gg}$ and $U_{ee}$ amplitudes do not have any strong implications on magnetic phases that can be approached with cold $^{173}$Yb atoms~\cite{Cichy2016PRA}.
Moreover, as we show below, the difference between $U_{gg}$ and $(V-V_{\rm ex})$ as well as $U_{ee}$, makes the system more susceptible to an orbital ordering instability, which also appears in transition metal compounds~\cite{Khomskii2014}.

For studying low-temperature phases, we employ DMFT, which is approximative and only exact in the limit of an infinite-dimensional system~\cite{Georges1996RMP}.
Nevertheless, it has become a well-accepted method successfully applied to strongly-correlated electron systems and has also found applications in the description of ultracold atoms in optical lattices~\cite{Schmidt13,Qin18,Sandholzer19}.
The calculations are performed with an exact-diagonalization impurity solver~\cite{Caffarel1994PRL}, which preserves the SU($2$) spin-rotational symmetry of the two-orbital Hubbard model in Eq.~\eqref{eq:hamiltonian}~\cite{Golubeva2017PRB}.
To measure observables in the symmetry-broken phases, we perform doubling of the unit cell, i.e., allow two different DMFT solutions on neighboring lattice sites.

\section{Results}\label{sec:results}
For the central case in our DMFT analysis, we choose a fixed set of Hubbard parameters, $t_e=0.26 t_g$, $U_{gg}=6.8 t_g$, $U_{ee}=17 t_g$, $V=32 t_g$, and $V_{\rm ex}=23 t_g$, corresponding to a typical quasi-2D SDL with the polarizability ratio $p=2.1$ (see Appendix~\ref{apx:hubbard-params}).
The exact choice of parameters is not crucial but motivated by the experimental accessibility and the signatures of orbital ordering considered in our study.
We briefly note that the interaction energies $V$ and $V_{\rm ex}$ are comparable to the band gap of the SDL and need to be renormalized accordingly.
Renormalizing these parameters is particularly nontrivial in our regime of quasi-2D geometry and mixed confinement.
Instead, we perform the renormalization on the basis of an approximative scheme and verify independently that our main results are not sensitive to the precise magnitude of $V$~and~$V_{\rm ex}$~(see~Appendix~\ref{apx:hubbard-params}).

\begin{figure}[t]
  \includegraphics[width=\linewidth]{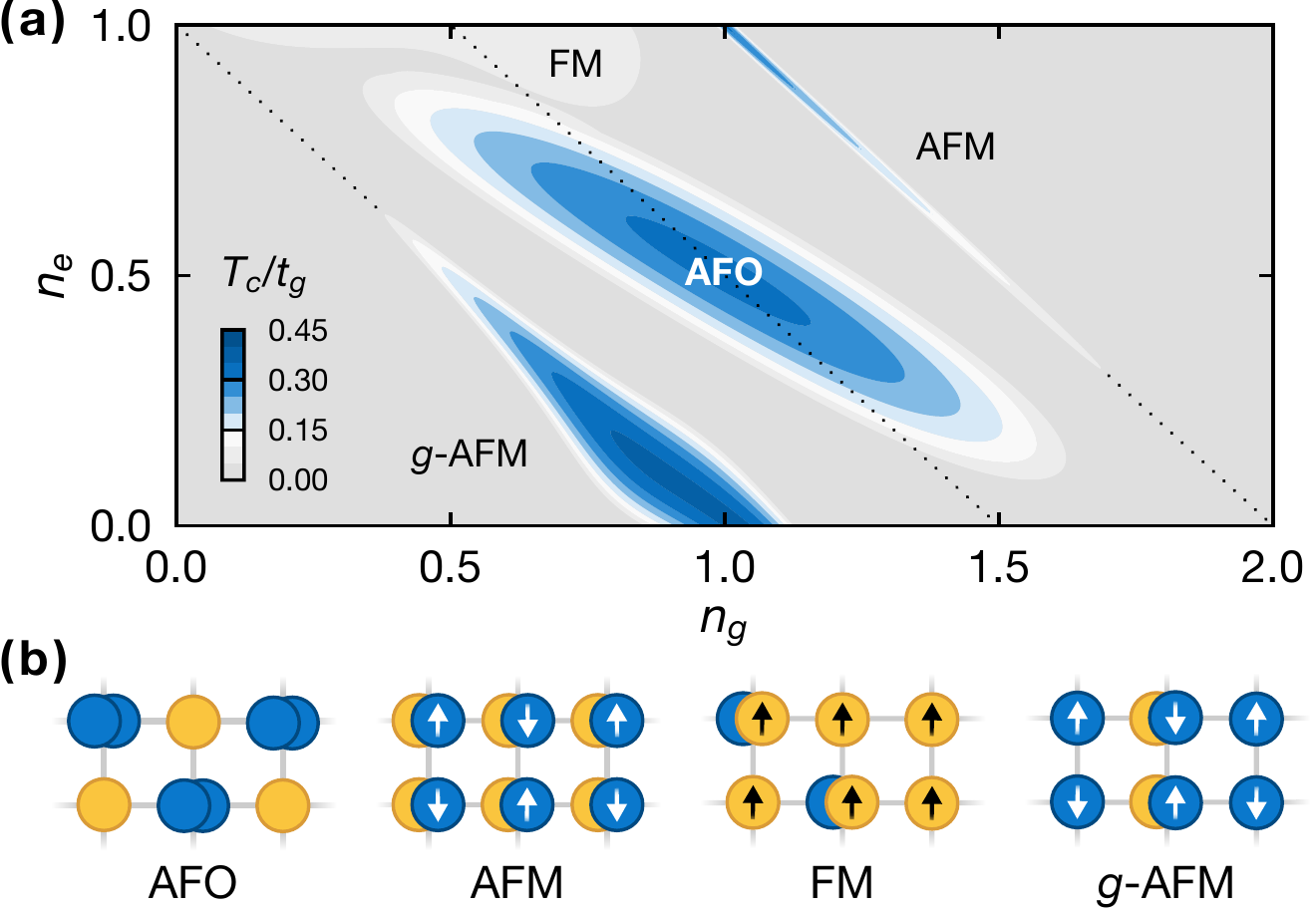}
  \caption{\label{fig:phase-diagram}%
    (a)~Phase diagram obtained from DMFT (see Appendix~\ref{apx:dmft}) for different densities $n_g$ and $n_e$ of $^{173}$Yb in a SDL with the polarizability ratio $p=2.1$.
    The filled contours indicate the critical temperatures $T_c$ of the ordered phases, which are antiferro-orbital (AFO), antiferromagnetic (AFM), ferromagnetic (FM), and antiferromagnetic in $g$ ($g$-AFM).
    The gray-shaded regions at \mbox{$T_{c}=0$} correspond either to a normal phase without long-range order or a regime with phase-separation and the dotted lines indicate constant total filling.
    (b)~Illustration of the local orbital and magnetic order in the different phases.
    Blue and yellow circles correspond to $g$ and $e$ atoms, arrows indicate the spin state, and lattice bonds are shown as gray lines.
    In the AFM and FM phases, the $g$ and $e$ spins are parallel on doubly-occupied lattice sites.}
\end{figure}

The low-temperature phase diagram derived from the DMFT calculation is shown in Fig.~\subfigref{phase-diagram}{a}.
For certain orbital fillings $(n_g,n_e)$ and at sufficiently low temperature $T_c/t_g \leq 1$ ($k_B = 1$ is used below), we distinguish ferromagnetic (FM), antiferromagnetic (AFM), and antiferro-orbital (AFO) long-range-ordered phases as illustrated in Fig.~\subfigref{phase-diagram}{b}.

Since magnetically-ordered phases are not the main focus of the current study, we only briefly comment on the key observations.
According to the diagram shown in Fig.~\subfigref{phase-diagram}{a}, the $g$-AFM and AFM instabilities appear along diagonal lines in the $n_g$-$n_e$ plane, where the total density $n$ is $\approx1$ or $\approx2$, respectively.
These are the Mott-insulating regimes with one exception at $n_g=2$, where the system becomes an insulator with vanishing local magnetic moments.
At $n_e\approx1$ and variable $n_g$, the Hamiltonian in Eq.~\eqref{eq:hamiltonian} can be mapped to the double-exchange model (FM Kondo-lattice model) with both FM and AFM terms~\cite{sotnikov18}.
We note that away from the polarizability ratio $p=1$, FM ordering becomes stabilized with increasing $p$ due to the larger exchange interaction~$V_{\rm ex}$ and the stronger localization of $e$ atoms.
In contrast, the AFM phases involving the $e$ orbital become suppressed due to the strong localization of $e$ atoms and an increase of the local interaction amplitudes.

The AFO phase is characterized by the alternating occupation of neighboring lattice sites with atoms in different orbitals.
In analogy to the N\'{e}el order of spins, the lattice can be viewed as a set of two sublattices: One is dominantly occupied by pairs of $g$ atoms, while the other is dominantly occupied by single atoms in the $e$ state.
This configuration is similar to orbital ordering in solid-state materials, where sublattices are formed by electrons occupying different orbital angular-momentum states of the lattice ions~\cite{Khomskii2014}.
In contrast to real crystals, our proposed implementation does not introduce directional or interorbital hopping, which makes it more feasible for direct experimental realizations.

A peculiar feature of the system under study is that the AFO phase is stabilized in a wide region around $n_g=1$ and $n_e=0.5$ as can be seen in Fig.~\subfigref{phase-diagram}{a}.
In most regions, this phase is accompanied by charge order, i.e., the periodic modulation of the total density~$n$.
The transition to the long-range-ordered state close to $n_g=1$ and $n_e=0.5$ is mainly driven by an interplay between the direct and superexchange interaction amplitudes.
This can be intuitively understood in the strong coupling limit, $t_e \ll t_g \ll U_{\rm eff} = (V - V_{\rm ex} - U_{gg})$.
In this case, the dominant superexchange amplitude~$\sim t_g^2/U_{\rm eff}$ reduces the total energy in an arrangement of pairs of $g$ atoms next to $e$ atoms on neighboring lattice sites, which yields the antiferro-orbital order illustrated in Fig.~\subfigref{phase-diagram}{b}.
We note that the system is actually in a slightly different regime with intermediate coupling ($t_g \lesssim U_{\rm eff}$), for which there is no exact analytical formula of the corresponding amplitude, but the intuitive picture remains valid.
Furthermore, it is worth mentioning that the AFO phase is not limited to the chosen 2D geometry.
Supporting calculations performed for both a three-dimensional~\cite{dmft-afo3D} and a one-dimensional system (matrix product state algorithm~\cite{Rizzi2018}), reveal qualitatively similar correlations in the orbital domain at comparable densities.

For our chosen set of Hubbard parameters, the AFO phase is enhanced compared to most magnetically-ordered phases and the DMFT analysis yields a transition point at the critical temperature $T_c/t_g = 0.31$.
We also obtain characteristic values for the entropy per particle required for orbital ordering, which are calculated from the Maxwell relation for the local density of atoms (see Appendix~\ref{apx:dmft}).
We estimate the maximal entropy per particle in the bulk for the AFO phase to be $s \approx 0.8$, which is related to approximately a tenth of the Fermi temperature in a harmonic trap under the assumption that loading into the optical lattice is adiabatic~\cite{Carr2004PRL,Koehl2006,Sotnikov2016PLA}.
Degenerate Fermi gases of $^{173}$Yb atoms in the $g$ state have been reported in a similar entropy and temperature regime~\cite{Hofrichter16PRX}, but the preparation of two-orbital mixtures is more challenging, since $g$ atoms need to be partly excited into the $e$ state for the desired orbital population (see Appendix~\ref{apx:implementation}).

In Fig.~\ref{fig:crit-quant}, we explore the influence of the polarizability ratio~$p$ on the critical temperature and entropy of the AFO phase.
\begin{figure}
  \includegraphics[width=\linewidth]{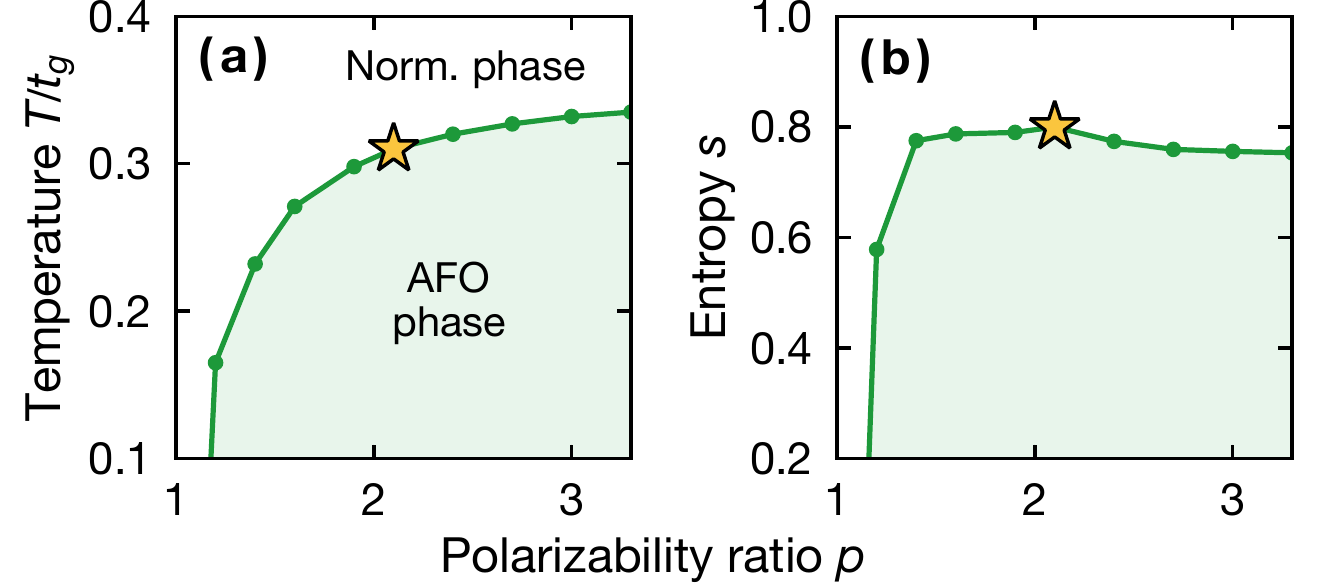}
  \caption{\label{fig:crit-quant}%
    (a)~Critical temperature and (b)~critical entropy per particle of the AFO phase for variable polarizability ratio $p$ at fixed densities $n_g=1$ and $n_e=0.5$.
    The yellow star indicates $p=2.1$, which is the central value of our study.
    Green circles refer to values obtained from DMFT and lines serve as a guide to the eye.}
\end{figure}
While the orbitally-ordered phase vanishes completely in the vicinity of $p=1$, it is stabilized with increasing $p$ and the critical temperature only changes negligibly for $p>2$.
Similarly, the critical entropy increases at small $p$, reaches its maximal value at moderate coupling ($p\approx2$), and then slowly decreases due to a stronger suppression of particle-number fluctuations at larger values of the interaction strengths.
This critical behavior is analogous to the one observed in the proximity of AFM phases~\cite{Golubeva2017PRB} and motivates the choice of $p=2.1$ for our study.
 
In particular, the AFO phase can be probed experimentally since it covers a sizable fraction of the $n_g$-$n_e$ phase diagram, as can be seen in Fig.~\subfigref{phase-diagram}{a}.
Due to its relative stability against particle-number fluctuations (thermally-induced metal-insulator crossover region at $n=1.5$) and almost equidistant separation from other insulating regimes ($n=1$, $n=2$, and $n_e=1$), it should only require relatively coarse tuning of the respective densities.

\begin{figure}[t]
  \includegraphics[width=\linewidth]{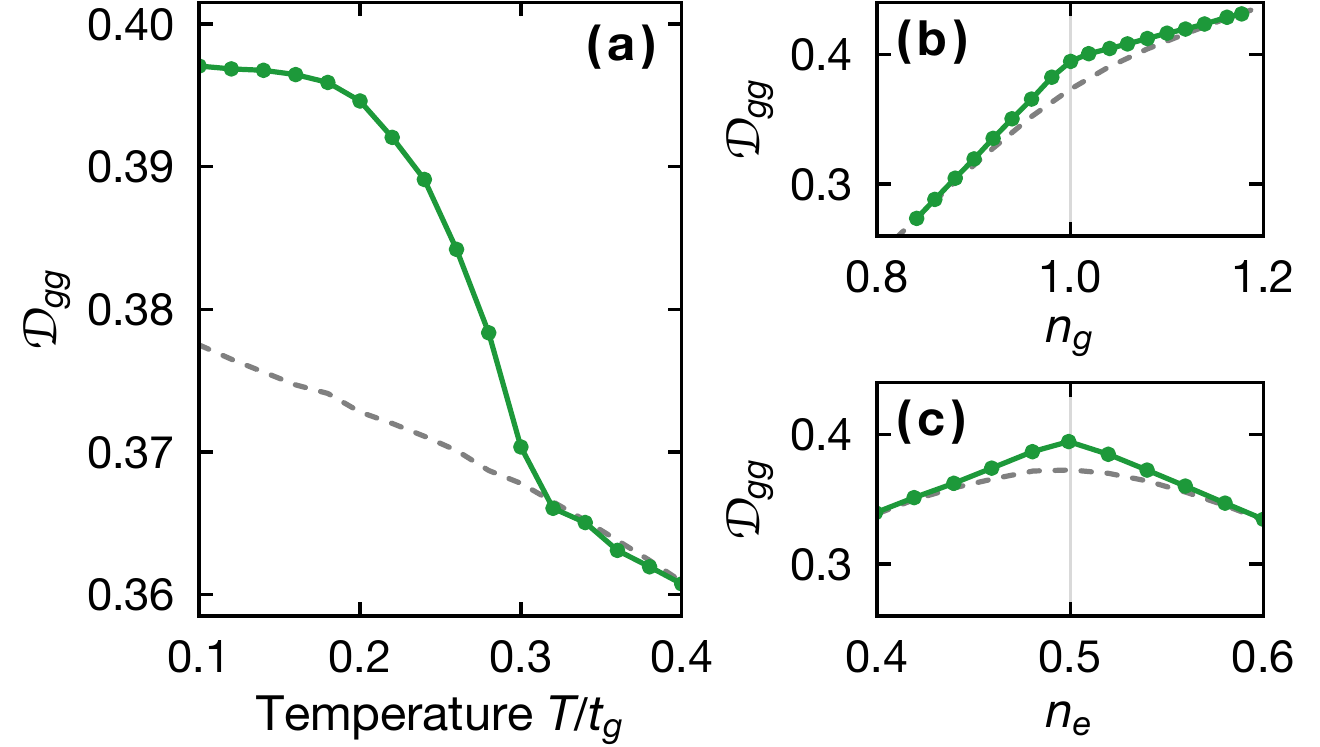}
  \caption{\label{fig:glob-observables}%
    Site-averaged double occupancy $\mathcal{D}_{gg}$ (a)~as a function of temperature at $n_g = 1$, $n_e = 0.5$ and as a function of the orbital density (b)~$n_g$ and (c)~$n_e$ at fixed $T/t_g = 0.2$.
    We show the result for the AFO phase in green and for an artificially-restricted normal phase in dark gray.
    Green circles refer to values obtained from DMFT and lines serve as a guide to the eye.}
\end{figure}

First, we analyze how the AFO ordering can be detected by measuring the fraction of lattice sites occupied by pairs of $g$ atoms, $\mathcal{D}_{gg}= \langle n_{ig\uparrow}n_{ig\downarrow} \rangle$.
This observable can be probed experimentally by measuring the $g$ atom number upon removal of atoms on doubly-occupied lattice sites with a resonant photoassociation pulse, a well-established measurement technique, which has been successfully applied to ultracold $^{173}$Yb atoms in optical lattices~\cite{Tai2012Nat}.
As shown in Fig.~\subfigref{glob-observables}{a}, we first keep the atomic densities fixed, $n_g = 1.0$ and $n_e = 0.5$, and vary the temperature.
In addition, we also study the $\mathcal{D}_{gg}$ dependencies at fixed temperature but variable $n_g$ or $n_e$ to quantify the sensitivity on the density in each orbital [see Figs.~\subfigref{glob-observables}{b}~and~\subfigref{glob-observables}{c}].
The temperature and density dependencies of $\mathcal{D}_{gg}$ clearly indicate the enhancement of local pairing of $g$ atoms in the AFO phase.
Below the critical temperature, this effect increases but approaches a saturated regime for temperatures $T/t_g \leq 0.2$.
The site-averaged $\mathcal{D}_{gg}$ signal in the AFO phase at $T/t_g=0.2$ differs from the one obtained in an artificially-restricted normal phase at the same temperature by $\approx8\%$.
Additional calculations with variable Hubbard parameters show that this value can be slightly increased by a reduction of the lattice depth in any of the three spatial directions.

\begin{figure}[t]
  \includegraphics[width=\linewidth]{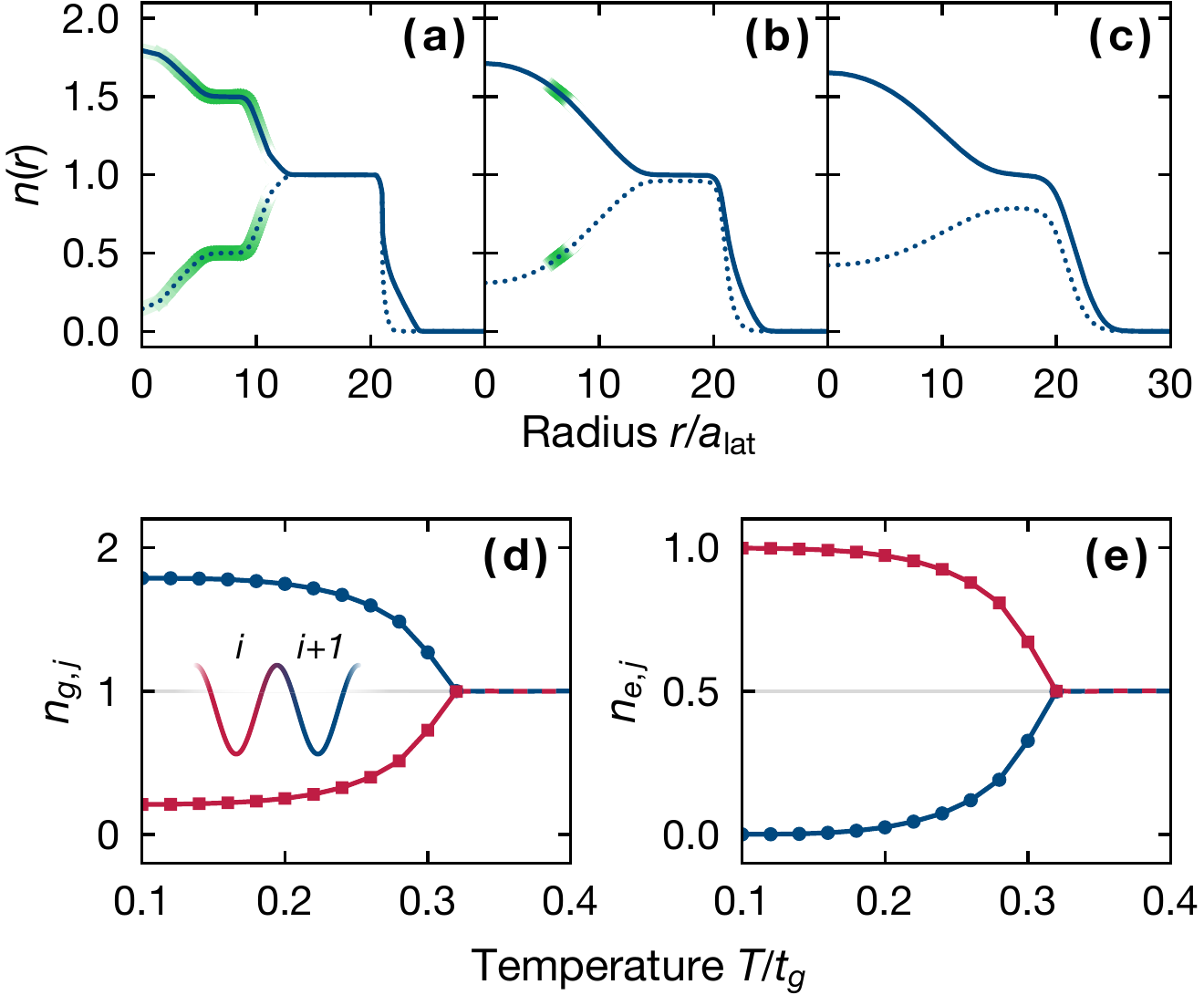}
  \caption{\label{fig:loc-observables}%
    \mbox{(a)--(c)}~Radial density profiles in a harmonic trap with the potential $V_{\rm ho}/t_g = 2.3 \times 10^{-2} ( r/a_{\rm lat} )^2$, which corresponds to the trapping frequency $\omega = 2\pi \times 60\asciimathunit{Hz}$.
    Here, $a_{\rm lat}=\lambda/2$ is the lattice constant.
    We show the profiles for fixed atom number $N\approx 1.7\times 10^{3}$ and temperatures (a)~$T/t_g=0.1$, (b)~$0.3$ (slightly below $T_c$), as well as (c)~$0.6$.
    Solid lines refer to the total density $n = n_e + n_g$ while dotted lines show $n_e$.
    The amplitude of the charge-density wave, $(n_{i}-n_{i+1})/(n_{i}+n_{i+1})$, indicates the orbitally-ordered region of the trap and is shown as thick green line in the background.      
    \mbox{(d),(e)}~Local density of the two orbitals on neighboring lattice sites, $j=i$~(red squares) and $j=i+1$~(blue circles), as a function of temperature at mean density $n_g = 1$ and $n_e = 0.5$.
    The solid lines serve as a guide to the eye.}
\end{figure}

In comparison to the global observable $\mathcal{D}_{gg}$, the AFO phase shows much stronger signatures in local quantities such as the in-trap density distribution, which can be measured directly with high resolution {\it in situ} imaging.
Ultracold atoms trapped in an attractive optical lattice potential usually experience harmonic confinement due to the curvature of the Gaussian laser beams.
The resulting smooth change of the chemical potential can then lead to the coexistence of multiple phases in a single trap, such as the well-known shell structure consisting of spatially-alternating Mott-insulating and metallic regions.
We explore the density profiles in a harmonic trap across the thermally-induced AFO phase transition in Figs.~\subfigrefs{loc-observables}{a}{c}.
Below the critical temperature, in parallel with the development of AFO correlations across the trap, we observe the formation of a Mott-insulating plateau at $n=1.5$, which is clearly visible in Fig.~\subfigref{loc-observables}{a}.
Interestingly, the phase separation of $g$ and $e$ atoms, as well as another Mott-insulating plateau at $n=1$, can already be observed above the critical temperature due to a decoupling from the superexchange energy scales [see Fig.~\subfigref{loc-observables}{c}].
This could allow detecting this signature as a precursor of the AFO phase in an experiment, even above the actual transition point~($T/t_g \lesssim 0.8$).

Next, we focus on density correlations between individual lattice sites, which could be directly probed with single-site resolved imaging of $g$ and $e$ atoms~\cite{Gross17}.
In Figs.~\subfigref{loc-observables}{d}~and~\subfigref{loc-observables}{e}, we show temperature dependencies of the densities $n_g$ and $n_e$ on two neighboring lattice sites across the AFO phase transition.
In general, these show strong signals from pair formation and redistribution of atoms in different orbitals on a checkerboardlike pattern in the lattice [see Fig.~\subfigref{phase-diagram}{b}].
Already slightly below $T_c$, the density $n_g$ ($n_e$) reaches $1.5$ ($0.25$) on the first site and $0.5$ ($0.75$) on the neighboring site.
We also expect the build-up of spatial correlations beyond nearest-neighbor sites, whose amplitudes cannot be accurately calculated within DMFT but could be probed in the experiment.
With recent advances in quantum gas microscopy of AEAs~\cite{Yamamoto16,Miranda17}, local density correlations could provide a direct detection method of the AFO phase and its properties.
Moreover, local control in these experiments could allow one to precisely engineer and study excitations in the AFO regime~\cite{Khomskii2014}.

\begin{figure}[t]
  \includegraphics[width=\linewidth]{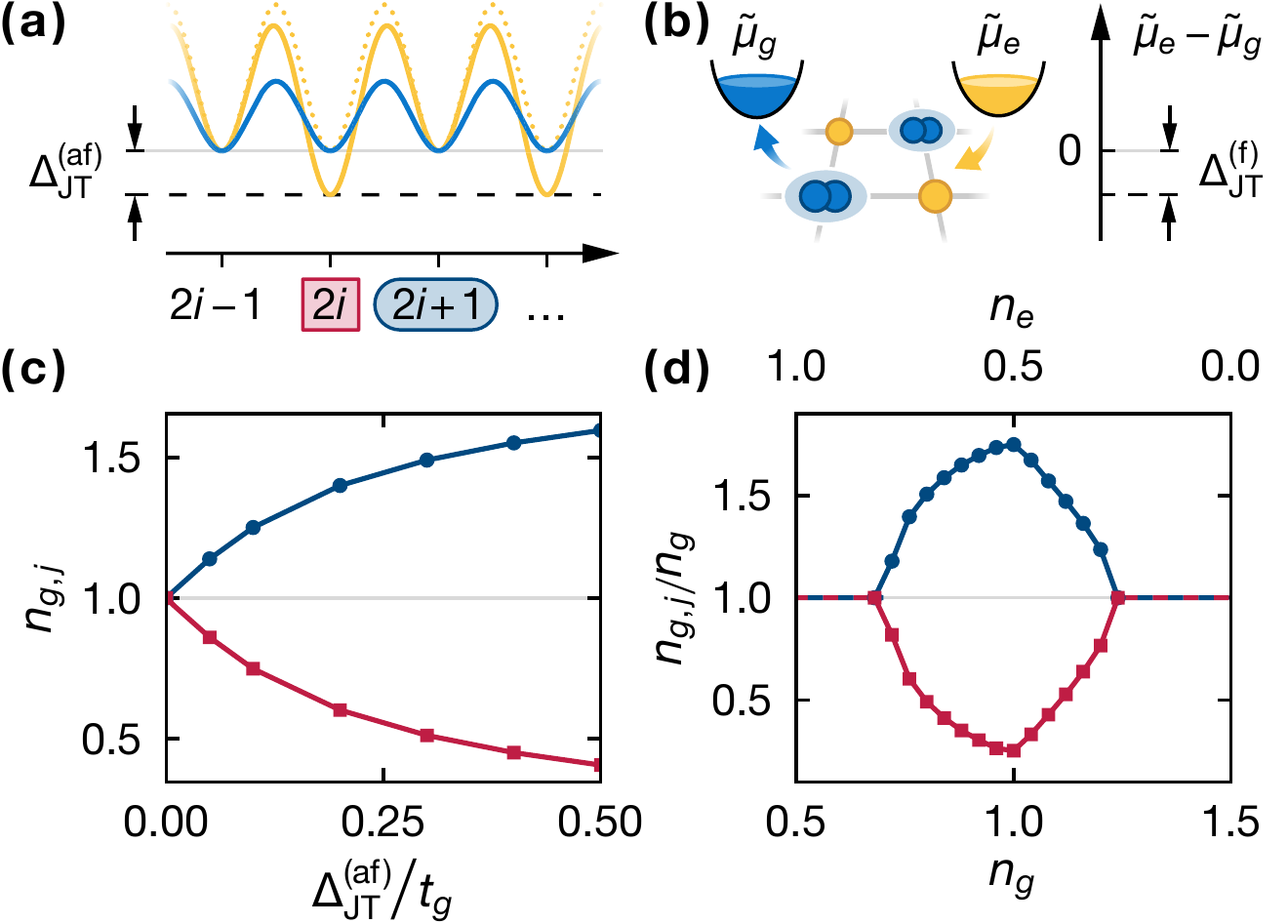}
  \caption{\label{fig:jte}%
    (a)~Schematic representation of the lattice potentials for $e$~(yellow lines) and $g$~atoms (blue lines) in the presence (solid) and absence (dotted) of the additional superlattice potential producing an offset~$\Delta_{\rm JT}^{\rm (af)}$ in analogy to the antiferrodistortive JTE\@.
    (b)~Illustration of changing the orbital densities at constant $n=1.5$ in the AFO phase and the resulting ferrodistortive JTE analog quantified by $\Delta_{\rm JT}^{\rm (f)}$.
    We show the limit of strongly-bound $g$ atoms (blue circles) and single $e$ (yellow circles) atoms, and their renormalized chemical potentials $\tilde{\mu}_g$ and~$\tilde{\mu}_e$.
    \mbox{(c),(d)}~Local (normalized) density of $g$ atoms on neighboring lattice sites, $j=2i$ (red squares) and $j=2i+1$ (blue circles), for (c)~probing the antiferrodistortive JTE with a variable superlattice potential at $n_g=1$, $n_e=0.5$, and $T/t_g=0.36$ and (d)~probing the ferrodistortive JTE at constant $n=n_e + n_g = 1.5$ and variable $n_g/n_e$ at $T/t_g=0.2$.
    The solid lines serve as a guide to the eye.
  }
\end{figure}

Finally, we discuss a potential extension of our proposal to host not only the superexchange-driven mechanism for orbital ordering but also include a source field comparable to the Jahn-Teller effect in transition metal oxides~\cite{Khomskii2014}.
For the analog of the antiferrodistortive JTE (as in K$_2$CuF$_4$)~\cite{Fazekas1999}, we consider adding a superlattice structure for the $e$ orbital as shown in Fig.~\subfigref{jte}{a}.
For the proposed $^{173}$Yb system, an additional optical lattice with the wavelength $\lambda \approx 1380\nm$ and $|p|\gg 1$ would produce a suitable superlattice potential, which acts predominantly on $e$ atoms and lowers their energy at every second site of the original SDL.
In our limit of a weak superlattice potential, Eq.~\eqref{eq:hamiltonian} acquires only an additional onsite term for $e$ atoms, $\mathcal{H}'=-\sum_{j=2i}\Delta_{\rm JT}^{\rm (af)}n_{e,j}$.
The DMFT analysis for this staggered potential confirms that the AFO phase can be substantially extended to higher temperatures with a transformation of the second-order transition point to a crossover regime due to the explicit symmetry breaking by the superlattice.
In Fig.~\subfigref{jte}{c}, we show the $g$ atom density on neighboring lattice sites at $T = 0.36\,t_g > T_c$, where a signal of the AFO phase [analogous to Fig.~\subfigref{loc-observables}{c}] emerges with increasing $\Delta_{\rm JT}^{\rm (af)}$.
In contrast, the analog of the ferrodistortive JTE (as in La$_2$CuO$_4$)~\cite{Fazekas1999} could be probed without additional potentials.
Its destructive impact on staggered orbital ordering can be analyzed by varying $n_g$/$n_e$ along a line of constant total density, in particular, $n=1.5$, shown as dotted line in Fig.~\subfigref{phase-diagram}{a}.
This effect can be intuitively understood for strongly-bound pairs of $g$ atoms and single $e$ atoms ($n_g=1$ and $n_e=0.5$).
In this limit, we can attribute the renormalized chemical potentials, $\tilde{\mu}_g$ and $\tilde{\mu}_e$, to the (compound) particles as illustrated in Fig.~\subfigref{jte}{b}.
Adjusting the average orbital densities such that $n_g/n_e \neq 2$
corresponds then to lifting the degeneracy of $\tilde{\mu}_g$ and $\tilde{\mu}_e$, which introduces an effective site-independent and thus ferrodistortive offset $\Delta_{\rm JT}^{\rm (f)}$.
In Fig.~\subfigref{jte}{d}, we plot the normalized density $n_g$ on neighboring lattices sites at $T =0.2\,t_g < T_c$, which reveals how the signatures of the AFO phase are suppressed by the change of $n_g/n_e$.
In general, the local observables for exploring the JTE analogues shown in Fig.~\ref{fig:jte} could be directly probed in the experiment by measuring correlations on neighboring lattice sites for variable superlattice depth or atomic densities.
Furthermore, we expect the global fraction of doubly-occupied sites $\mathcal{D}_{gg}$ to also show similar but less pronounced signatures.

\section{Summary and Outlook}\label{sec:summary-outlook}
We show that AEAs in SDLs are promising candidates for the experimental observations of orbital ordering phenomena and potentially could improve the understanding of related mechanisms in solid-state materials.
In particular, by means of changing the lattice depth and polarizability ratio between different orbital states, a capability to enhance or suppress the superexchange contributions to the AFO ordering instability is demonstrated.
At the same time, in a well-controlled and independent manner, contributions analogous to the JTE in crystals could be explored by adjusting the orbital densities or by introducing a superlattice potential.
The rich structure of the phase diagram revealed in this study also makes AEAs in SDLs suitable for studies of open questions on the critical behavior and excitations in transition-metal oxides hosting orbitally-ordered as well as various magnetic and superconducting phases~\cite{Tokura00,Khalifah02,Keimer06,Khomskii2014,Singh15}.

Our analysis oriented towards experimental implementations with $^{173}$Yb atoms reveals that the SDL substantially increases the difference between the intraorbital interactions, $(U_{ee}-U_{gg})\gtrsim U_{gg}$.
Therefore, the AFO instability crucially depends on the energy gap to the closest interorbital excitation [$(V-V_{\rm ex}-U_{gg})$ for $V_{\rm ex}>0$].
This small gap gives the largest contribution to the corresponding AFO superexchange amplitude, which depends less on $U_{ee}$ and the energy of the other interorbital excitation [$(V+V_{\rm ex}-U_{gg})$  for $V_{\rm ex}>0$].
Therefore, similar calculations and experiments could be realized with related species, such as $^{87}$Sr or $^{171}$Yb.
While the former and $^{173}$Yb have comparable ordering of the interaction parameters~\cite{Goban2018}, the latter features antiferromagnetic exchange interaction $V_{\rm ex} < 0$~\cite{Ono19} and almost vanishing $|U_{gg}|\ll t_g$~\cite{Kitagawa08}, which could provide an interesting extension of the phase diagram discussed in our study.

At higher spin symmetry, the AFO phases may demonstrate unconventional space modulations involving more than two sublattices.
These could naturally be studied with $^{173}$Yb when the large $\text{SU}(N\leq 6)$ symmetry in the $g$ and $e$ orbital is utilized~\cite{Gorshkov2010NP}.
Another related effect concerns the potential magnetic order of the SU($2$)-symmetric mixture in the AFO phase at very low temperatures, which requires a comprehensive analysis of potential sublattice structures and remains an interesting task for future theoretical research.

\begin{acknowledgments}
The authors thank Andreas Haller, Atsushi Hariki, Jan Kune\v{s}, Luis Riegger, Matteo Rizzi, and Sebastian Scherg for helpful discussions.
We also thank Vladimir A. Dzuba for providing the polarizability values from Ref.~\cite{dzuba18}.
A.S. acknowledges support by the European Research Council (ERC) under the European Union's Horizon 2020 research and innovation program (Grant Agreement No. 646807-EXMAG)
and by the Ministry of Education and Science of Ukraine (Research Grant with internal University No.~07-13-20). 
N.D.O. acknowledges funding from the International Max Planck Research School for Quantum Science and Technology.
Y.Z. and A.C. acknowledge funding of this work by the National Science Centre (NCN, Poland) under Grant No. UMO-2017/24/C/ST3/00357.
Access to computing and storage facilities provided by the Poznan Supercomputing and Networking Center (EAGLE cluster) is greatly appreciated.
\end{acknowledgments}

\appendix

\section{Experimental implementation}\label{apx:implementation}
Realization of the orbitally-ordered phase discussed in the main text requires the preparation of $g$ and $e$ atoms at variable density in a 2D SDL at low enough temperatures.
We first focus on the optical lattice implementation and briefly outline possible state preparation techniques.

For $^{173}$Yb, a monochromatic SDL at a wavelength of $670\nm$ (polarizability ratio $p=3.3$) has been implemented in one dimension~\cite{Riegger18} and can be realized similarly in 2D.
For our choice of $p=2.1$, theoretical calculations of the polarizability~\cite{dzuba18} yield a wavelength of $690\nm$, which is accessible with commercial laser systems.
We note that the precise value of this wavelength has only negligible influence on the results discussed in the main text.
For the SDL, we consider a fixed lattice depth of $V_{x,y}=5E_{\rm rec}^{\rm SDL}$ ($g$ atoms) to ensure strong suppression of next-nearest-neighbor tunneling and the validity of the tight-binding approximation.
For the strong confinement along $z$, we consider a deep magic-wavelength ($\lambda = 759\nm$) lattice, $V_z=18\,E_{\rm rec}^{m}$, such that the system is in the quasi-2D regime.
Here, $E_{\rm rec}^{\rm SDL}=h\times2.4\asciimathunit{kHz}$ and $E_{\rm rec}^m=h\times2.0\asciimathunit{kHz}$ refer to the recoil energy from a photon of the SDL or magic-wavelength lattice, respectively.

The two-orbital mixture can be prepared in the optical lattice by optically exciting part of the $g$ atoms with an appropriate laser pulse~\cite{Riegger18}.
Besides the orbital degree of freedom, $^{173}$Yb atoms feature six nuclear spin states in $g$ and $e$, with $m_F\in\{-5/2,-3/2,\ldots,+5/2\}$.
Due to SU($N$)-symmetric collisions, a stable subset of these states can be prepared and used in the experiment~\cite{Gorshkov2010NP,Scazza2014NP}.
For the realization of the Hamiltonian in Eq.~\eqref{eq:hamiltonian}, we only consider two spin states, $m_F=-5/2$ and $+5/2$ (denoted by $\downarrow$ and $\uparrow$), as discussed in the main text.
For the state preparation, we suggest utilizing two additional spin states, $m_F=-3/2$ and $+3/2$ (denoted by $\swarrow$ and $\nearrow$).
Optical pumping on the intercombination line allows preparing $g$ atoms in an imbalanced mixture of all four spin states such that $(n_\swarrow + n_\nearrow)/(n_\downarrow + n_\uparrow)$ equals the desired ratio of $n_e/n_g$.
Subsequent transfer of the ancillary states ($\swarrow$ and $\nearrow$) into the $e$ orbital with circularly-polarized light yields the desired densities $n_g$ and $n_e$ with spin states $\downarrow$ and $\uparrow$ (see Ref.~\cite{DarkwahOppong2019} for a similar technique).

\section{Hubbard parameters}\label{apx:hubbard-params}
We calculate the Hubbard parameters from the numerical solution of the band structure of a separable three-dimensional optical lattice in the tight-binding approximation and with the corresponding lattice depths discussed in Appendix~\ref{apx:implementation}.
Since $g$ and $e$ atoms experience different lattice depths in the $(x,y)$ plane and along $z$, we use independent band structures for each orbital and spatial direction.
We list all relevant parameters in Table~\ref{tbl:hubbard-params} for a range of polarizability ratios considered in the main text.
\begin{table}[t!]
  \begin{ruledtabular}
    \caption{\label{tbl:hubbard-params}%
      Hubbard parameters for three different polarizability ratios and SDL wavelengths~\cite{dzuba18,Riegger18} at fixed lattice depths $(V_{x,y},V_z) = (5E_{\rm rec}^{\rm SDL}, 18\,E_{\rm rec}^{m})$.
      The column in bold font indicates the central values of our study ($p=2.1$, $\lambda=690\nm$).
      All parameters are given in units of the tunneling amplitude $t_g$ unless noted otherwise.
      The quantities $U_{eg}^+$, $V$, and $V_{\rm ex}$ are renormalized, while the values in brackets are directly obtained from Eq.~\eqref{eqn:onsite-interaction}.}
    \begin{tabular}{l @{\extracolsep{\fill}} c @{\quad} c @{\quad} c }
        Polarizability ratio $p$ & 3.3 & {\bf 2.1} & 1.2\\
        SDL wavelength $\lambda$ (nm) & 670 & {\bf 690} & 730 \\[0.35em]
        \colrule\\[-1.2em]
        $t_g$ ($h\times\mathrm{Hz}$) & 170  & {\bf 160}  & 143 \\
        $t_e$ & 0.07  & {\bf 0.26}  & 0.77 \\
        $U_{gg}$ & 6.78  & {\bf 6.78}  & 6.78 \\
        $U_{ee}$ & 22.3  & {\bf 17.0}  & 11.8 \\
        $U_{eg}^-$ & 10.3  & {\bf 9.33}  & 7.95 \\
        $U_{eg}^+$ & 60.1 [88.0]  & {\bf 55.0} [79.6]  & 49.7 [67.9] \\
        $V = (U_{eg}^+ + U_{eg}^-)/2$ & 35.2 [49.2]  & {\bf 32.2} [44.5] & 28.8 [37.9] \\
        $V_{\rm ex} = (U_{eg}^+ - U_{eg}^-)/2$ & 24.9 [38.8]  & {\bf 22.9} [35.1] & 20.9 [30.0] \\
    \end{tabular}
  \end{ruledtabular}
\end{table}

The onsite interaction strength $U_{\gamma\gamma'}$ is typically calculated from the corresponding s-wave scattering length $a_{\gamma\gamma'}$,
\begin{align}
    U_{\gamma\gamma'} = \frac{4\pi\hbar^2}{m} a_{\gamma\gamma'} \int \!{\rm d}^3 r\; w_{\gamma}^2(\mathbf{r}) w_{\gamma'}^2(\mathbf{r}).\label{eqn:onsite-interaction}
\end{align}
Here, $m$ is the atomic mass and $w_{\gamma}(\mathbf{r})$ is the Wannier function of the corresponding orbital $\gamma \in \{g, e\}$ derived from the band-structure calculation.
In the limit of large scattering lengths comparable to the lattice spacing, $a \sim a_{\rm lat}$, contributions from higher bands of the optical lattice become sizable.
Nevertheless, such a system can still be described within the lowest-band approximation
by absorbing these contributions into renormalized Hubbard parameters~\cite{Buechler10,Luhmann2012}.
In our case, $a_\mathrm{lat} \approx 6 \times10^3 a_0$ is the (smallest) lattice constant with $a_0$ the Bohr radius.
For the intraorbital scattering lengths $a_{gg}= 199 a_0$~\cite{Kitagawa08}, $a_{ee}=306 a_0$~\cite{Scazza2014NP}, and the interorbital singlet scattering length $a_{eg}^-= 220 a_0$~\cite{Scazza2014NP,Hoefer15}, the corrections are small and neglected.
However, the large orbitally-symmetric scattering length $a_{eg}^+ \approx 2 \times10^3 a_0$~\cite{Hoefer15} leads to a significant correction of the corresponding amplitude $U_{eg}^+$, which would otherwise exceed the band gap.

The system discussed in the main text features anisotropic and mixed confinement due to the SDL and the quasi-2D geometry, which prevents us from directly applying existing results for the renormalization of $U_{eg}^+$~\cite{Buechler10}.
Instead, we use the geometric mean of both orbitals as the effective lattice depth and approximate each lattice site with a harmonic oscillator potential~\cite{Busch98,Riegger18}.
In addition, we apply first-order perturbation theory to account for the anharmonic cosine potential of the optical lattice.
Finally, we assume spatial separability of the problem and calculate two independent solutions for the $(x,y)$ as well $z$ direction, which we combine into the single interaction amplitude $U = U_{x,y}^{2/3}U_z^{1/3}$.
When applied to an isotropic system with comparable lattice depths, our results reasonably agree with Ref.~\cite{Buechler10}.
We find an onsite interaction energy $U_{eg}^+$ in excess of the band gap of $g$ atoms along $x$ and $y$ by up to $60\%$ ($p=3.3$), which suggests that our approximate approach fails to correctly predict the renormalized Hubbard parameter.
Although the effective $U_{eg}^+$ in the experiment will be different, we verify that the phases discussed in the main text are robust against variation of this parameter on a similar scale.

The large scattering length $a_{eg}^+$ also causes an increased relevance of non-Hubbard terms in the Hamiltonian, specifically, direct off-site interactions and density-assisted tunneling~\cite{Luhmann2012}.
While we expect the former to be negligible in our regime, the latter could become comparable to $t_g$ for the orbitally-symmetric interaction channel.
We cannot directly incorporate this term into our DMFT calculation, but the main effect will be a renormalization of the hopping amplitudes for sites occupied simultaneously by $g$ and $e$ atoms.
In principle, these excitations should mainly occur virtually in the AFO phase at $n_g\leq1.0$ and $n_e\leq0.5$.
At higher densities, we assume the effects can be absorbed into a modified $U_{eg}^+$, respectively $V$ and $V_{\rm ex}$, which again should not alter the phase diagram significantly.

\section{DMFT calculation}\label{apx:dmft}
In the DMFT analysis, we employ an exact diagonalization solver for the Anderson impurity problem with up to four bath orbitals per each spin and orbital component.
The DMFT self-consistency conditions for two sublattices are applied in the analysis of the AFO and AFM phases, while the normal and FM phases are analyzed within the single-site lattice projection~\cite{Cichy2016PRA}.

We obtain the inhomogeneous distributions in the harmonic trap and entropy dependencies within the local density approximation.
The entropy is calculated by numerical integration of the Maxwell relation, $S=\int\!{\rm d}\mu\,(\partial n/\partial T) $ on the interval from the vacuum state, $S(\mu_g^0,\mu_e^0)=0$, to the chemical potential values $\mu_{g}$ and $\mu_{e}$, which yield the desired densities of atoms, $n_{g}=1$ and $n_{e}=0.5$, in particular.

For the phase diagram in Fig.~\ref{fig:phase-diagram}, we fit the parameters $\theta$, $\mathbf{n}_0 = (n_{g,0}, n_{e,0})$, and $p_{ij}$ of the polynomial function,
\begin{align}
    f_{T_c}(\mathbf{n}) = \sum_{i, j = 0}^{N=2} p_{ij} {\left[R(\theta) \left(\mathbf{n}-\mathbf{n}_0\right)\right]}_g^i {\left[R(\theta) \left(\mathbf{n}-\mathbf{n}_0\right)\right]}_e^j
\end{align}
to a dense enough mesh of DMFT data points $(\mathbf{n}, T_c) = [(n_g,n_e),T_c]$ for each phase individually and evaluate this function in an appropriate region.
Here, $R(\theta)$ is the matrix, which rotates points through the azimuth angle $\theta$.

\bibliography{yb173_afo}

\end{document}